# Inherent Heating Instability of Direct Microwave Sintering Process: Sample Analysis for Porous 3Y-ZrO$_2$

Charles Manière[a], Tony Zahrah[b], Eugene A. Olevsky[a, c]*

(a) Powder Technology Laboratory, San Diego State University, San Diego, USA
(b) Matsys Inc., Sterling, USA
(c) NanoEngineering, University of California, San Diego, La Jolla, USA

**Abstract**

Direct microwave heating of 3Y-ZrO$_2$ is studied at frequency of 2.45 GHz. Different conditions of input power, sample position and size are tested. For the first time, the experimentally known instability of microwave sintering is explained coupling the effective medium approximation and finite-element method. We show how the material dielectric permittivity imaginary part which increases with temperature and relative density encourages high hot spot phenomena. It is shown that the sample location has a great impact on the temperature distribution and decreasing the sample size promotes temperature homogenization thereby assisting the overall sintering stabilization.

___________________________________

\* Corresponding author: **EO**: Powder Technology Laboratory, San Diego State University,

5500 Campanile Drive, San Diego, CA 92182-1323,

Ph.: (619)-594-6329; Fax: (619)-594-3599, *E-mail address*: eolevsky@mail.sdsu.edu

Zirconia ceramics presents useful properties in terms of super-plasticity, strength, hardness, ionic conductivity, and low thermal conductivity [1-3]. These properties lead to many industrial applications such as machinery, dentistry, oxygen sensors, oxygen pumps, solid oxide fuel cells (SOFCs), thermal barrier coating and refractory ceramics [4-7]. Many consolidation processes are reported in the literature for zirconia using various fabrication methods including conventional sintering, hot pressing, spark plasma sintering, and hot isostatic pressing [8-9]. This material is also an excellent candidate for flash sintering because of its thermally activated ionic conductivity [10]. The dielectric losses of zirconia in microwave illumination have also interesting values at high temperatures [11]. The microwave sintering of zirconia is widely present in the research field and in industry, including the fabrication of artificial teeth in dentistry [12].

It is known that the microwave sintering technique allows consolidation of various materials with high heating rates (100 K/min or higher) based on volumetric heating of the sample; it promotes fine microstructures with high levels of densification [13]. Numerous researches reported a significant enhancement of the mass transport during microwave sintering compared to the conventional sintering for the same thermal cycle conditions [11, 14-15]. In the literature the two main possible explications for this so called "non-thermal effect" are diffusion mechanism enhancements [16] and/or ponderomotive force action at the pore/grain boundary interface junctions [17, 18].

"Direct heating" is applicable if the specimen is sensitive to microwaves. For the non-dissipative or microwave- transparent materials, an indirect heating is also applicable using a supporting material dubbed "susceptor", which is highly receptive to microwaves. Silicon carbide is often used as the material for suceptors due to SiC significant dielectric loss values at low and high temperatures and due to SiC high melting point. The microwave heating of zirconia often uses susceptors because of its low dielectric loss properties at low temperatures [19]. However, in the "computer model experiment" used in the present work, the 3Y-ZrO$_2$ samples (55% of relative density) are placed in a resonant TE$_{102}$ mono-mode waveguide. This mode is characterized by two maximums electric field areas where the sample is introduced to allow the sample direct microwave heating.

Non-magnetic dielectric ceramic materials with a complex relative dielectric permittivity $\varepsilon_r = \varepsilon'_r - j\varepsilon''_r$ have a power dissipation per unit volume given by:

$$P = \omega\varepsilon_0\varepsilon''_r \boldsymbol{E}^2 \tag{1}$$

where $\omega$ is the angular frequency, $\varepsilon_0$ is the vacuum permittivity and $\boldsymbol{E}$ is the electric field strength. Consequently, the power dissipation depends on the frequency, the electric field and the dielectric permittivity imaginary part, which itself depends on the electric conductivity, temperature and relative density. The temperature evolution of real and imaginary parts of $\varepsilon_r$ of 3Y-ZrO$_2$ can be found in Ref. [11]. To determine the evolution with the relative density, it is possible to employ the effective medium approximation (EMA). This approach has been used to determine the dielectric properties of porous mixtures of different materials and in core-shell systems, etc. [13]. We use here the following classical EMA expression to determine the relative density deviation of the dielectric permittivity at different temperatures:

$$C_s \left(\frac{\varepsilon_{3YZ}-\varepsilon_{eff}}{\varepsilon_{3YZ}+2\,\varepsilon_{eff}}\right) + (1-C_s)\left(\frac{\varepsilon_{gas}-\varepsilon_{eff}}{\varepsilon_{gas}+2\,\varepsilon_{eff}}\right) = 0 \tag{2}$$

with $\varepsilon_{3YZ}$ being zirconia complex dielectric permittivity, $\varepsilon_{gas}$ - gas dielectric permittivity equal 1, $\varepsilon_{eff}$ - effective complex dielectric permittivity of the whole porous material to be determined, and $C_s$ - the solid concentration associated with the relative density. The results of the calculations of the real and imaginary parts of 3Y-ZrO$_2$ dielectric permittivity are reported in Fig.1 and the respective material parameter values are given in Table 1 [11, 20]. The real part increases with the temperature and relative density and is about 10 (at 1250 K and 55% of relative density). The imaginary part is very low at low temperatures (between 273 – 673K) and is about 0.02 for that temperature range. These low values rapidly increase after 673 K up to the values of the order of magnitude of 30.

Based on the determined dielectric properties of 3Y-ZrO$_2$, the electromagnetic heating is investigated by the finite element method (FEM) using COMSOL Multiphysics$^{TM}$ software.

The electromagnetic wave propagation obeys the Maxwell's equations, based on which COMSOL's 'RF module' can simulate the electromagnetic wave propagation using the following expression:

$$\nabla \times (\mu_r^{-1} \nabla \times \boldsymbol{E_r}) = k_0^2 \left(\varepsilon_r - \frac{j\sigma}{\omega \varepsilon_0}\right) \boldsymbol{E_r} \qquad (3)$$

with $\mu_r$ being the relative permeability, $k_0$ - the vacuum wave number, $\sigma$ - the electric conductivity and $\boldsymbol{E_r}$ is defined by the harmonic electric field expression $\boldsymbol{E} = \boldsymbol{E_r} exp(j\omega t)$. The thermal part of the problem is described by the heat transfer and electromagnetic loss equations:

$$\rho C_p \frac{\partial T}{\partial t} + \nabla \cdot (-\kappa \nabla T) = Q_e \qquad (4)$$

$$Q_e = \frac{1}{2} Re(\boldsymbol{J}.\boldsymbol{E}^*) + \frac{1}{2} Re(j\omega \boldsymbol{B}.\boldsymbol{H}^*) \qquad (5)$$

where $\rho$ is the density, $C_p$ is the specific heat, $\kappa$ is the thermal conductivity, $\boldsymbol{J}$ is the current density, $\boldsymbol{B}$ **is** the magnetic flux density, and $\boldsymbol{H}$ is the magnetic field intensity.

For this study, as mentioned above, a resonant rectangular wave guide is analyzed in a $TE_{102}$ mode. A rectangle cross-section with dimensions of $a = 80$ and $b = 40$ mm is considered for the 3D wave guide (see Fig.2a). In order to obtain resonance, the wave guide length must be a multiple of half the wave guide wave length ($\frac{\lambda_{wg}}{2}$); hence, the equation (6) can be used (for rectangular section wave guides) [21].

$$\text{For } TE_{nmp} \text{ mode } \quad \omega_{nmp} = c\sqrt{\left(\frac{n\pi}{a}\right)^2 + \left(\frac{m\pi}{b}\right)^2 + \left(\frac{p\pi}{l}\right)^2} \qquad (6)$$

where $a$, $b$ are the rectangular cross-section dimensions, $l$ is the wave guide length to be determined, $c$ is the light velocity, and $\omega_{nmp}$ is the resonant frequency. The calculated wave guide length for the chosen rectangular cross-section dimensions is then $l = 190$ mm. The simulation results in the empty cavity reported in Fig.2b use the $TE_{10}$ excitation port condition at the frequency of 2.45 GHz in one of the rectangle face extremities and employ perfect reflecting conditions on the other faces. As expected, the two maximum electric field areas corresponding to the $TE_{102}$ mode appear precisely at a quarter and three quarters of the cavity length *l*. When a 3Y-ZrO$_2$ cylindrical sample with dimensions of 12 mm in diameter, 12 mm in height and 55% of the relative density is considered at 300 K, as shown in Fig.2c, the field repartition changes. The maximum electric field is found to be located at y = 145.7 mm. At 300 K the relative permittivity imaginary part is low but the high value of the real part tends to repel the electric field lines outside the sample. Consequently, the electric field shown in Fig.2c is higher outside the sample and lower inside it.

The direct microwave heating problem formulation considers the sample to be placed at mid-height in the waveguide and in vacuum. The sample is located on a thermal insulating and microwave transparent support of fibrous alumina-silicate. In order to model this configuration a thermal insulation is placed at the bottom sample surface and the other faces are assumed to be subjected to pure thermal radiation:

$$\varphi_r = \sigma_s \, \epsilon \, (T_e^4 - T_a^4) \tag{7}$$

where $\varphi_r$ is the radiative heat flux, $\sigma_s$ is the Stefan-Boltzmann's constant, $\epsilon$ is the emissivity (0.7 for zirconia [22]), $T_e$ is the emission surface temperature, $T_a$ is the ambient cavity temperature.

The calculated temperature curves, electric field intensity and electromagnetic losses are reported in Fig.3 for constant input power ranging from 500 to 2000 W. Each simulation cycle was stopped when the maximum temperature reached 1250 K, - the temperature level where the densification starts [11]. The temperature curves (Fig.3a) present the first stage of the low rate heating corresponding to the low $\varepsilon_r''$ values at low temperatures. At 673 K a rapid and inhomogeneous heating is observed corresponding to the transition area of $\varepsilon_r''$ where its values highly increase. For 500 W the temperature cannot reach 673 K and the microwave coupling activation is not present. The hot spot phenomenon shown in Figs.3b, 3c, 3d indicates temperature differences of the order of magnitude of 400 K even when the power input is decreased. The electromagnetic losses (Figs.3e, 3f, 3g) have a distribution closely resembling the one of the temperature field (Figs.3b, 3c, 3d) and reflect the input power magnitude difference. The vertical temperature distribution is not symmetric because a thermal insulation plate is placed on the bottom. Alternatively, it is possible to place an insulating plate on the upper face to make the vertical temperature distribution symmetric.

This hot spot phenomenon (heating and, in turn, sintering instability) has been very often reported in the literature [23-24] and we can now understand its origin. Considering the relative dielectric permittivity imaginary part map (Fig.1b) and equation (1), the power dissipation increases with $\varepsilon_r''$ that itself increases with temperature and relative density. Consequently, the more the temperature is, the higher is the local dissipation. A small temperature gradient appears during the first stage of heating because of the boundary heat radiation, then this temperature gradient tends to be magnified (instead of being balanced out) because of the temperature-based increase of $\varepsilon_r''$. The hot spot phenomenon may happen in all materials where $\varepsilon_r''$ increases with the temperature, which is typical for most of the dielectric ceramics ($ZrO_2$, SiC, $Al_2O_3$, ZnO, $WO_3$, etc). If both the densification and the grain growth are thermally activated, then the hot spot phenomenon leads to high densification/grain size in-homogeneities and significant sample shape distortions.

The modification of the temperature distribution of the specimen by changing the sample location along Y axis is shown in Figs.4a, 4b, 4c. A change of only few millimeters of the sample position moves the sample away from the electric field maximum point and thus modifies the hot spot location. The modification of the sample dimensions as shown in Figs.4d, 4e, 4f gives similar results. The amount of energy dissipated is higher in samples with larger dimensions. Consequently, the location of the maximum electric field in the cavity is modified, and then the hot spot location in the sample is shifted too. More importantly, this result shows that the smaller the sample dimensions are, the higher is the temperature homogeneity. On the opposite, large size samples show pronounced localized heating.

The EMA results shown in Fig. 1b indicate the $\varepsilon_r''$ increase with the relative density leading to an increase of the localized heating. Based on the results shown in Fig. 4, the reduction of the sample size promotes heating homogenization. However, since the temperature differences are still high in Fig. 4d, the hot spot should also appear during densification.

To conclude, the obtained results provide a basis for explaining the inherent instability and non-reproducibility of the experiments frequently reported in the literature for microwave sintering [25-27]. The process instabilities stem from the intrinsic non-homogeneous electro-magnetic fields repartition and their sensitivity to the presence of dissipative materials during microwave sintering. The EMA method allows the calculation of the relative density dependence of the relative permittivity. The imaginary part responsible for dielectric losses increases with temperature and relative density and promotes the formation of highly localized hot spot at the center of the sample. This temperature inhomogeneity can be balanced by the use of susceptors which can add an external heat flux uniformizing the temperature inside the sample. This study shows in addition that it is also possible to homogenize the temperatures using small size samples when utilizing microwave sintering.

**Figure captions**

Fig. 1:    Zirconia permittivity as function of temperature and relative density calculated by the effective medium approximation: a) real part, b) imaginary part.

Fig. 2:    a) Dimensions of the $TE_{102}$ resonant mode cavity, b) electric and magnetic field intensity inside the empty cavity, c) cavity with the zirconia sample.

Fig. 3:    Heating simulations of 3Y-ZrO2 at 55% of relative density: a) min, max, average temperature evolution for samples heated at different constant input powers; (b, c, d) sample temperature fields and electromagnetic losses for input power between 1000 and 2000 W at Tmax=1250 K (e, f, g).

Fig. 4:    Sample temperature fields at Tmax=1250 K for 2000 W: (a, b, c) with different sample locations, (d, e, f) with different sample dimensions.

**Table captions**

Table 1:    Material and process electromagnetic-thermal parameters used in FEM calculations

# Table

Table 1: Material and process electromagnetic-thermal parameters used in FEM calculations [11, 20]. *(one column wide)*

| | Temperature range (K) | Expression |
|---|---|---|
| $Cp$ (J.kg$^{-1}$.K$^{-1}$) | 273-1473 | (43+2.35 T-0.34E-3 T$^2$+4.25E-6 T$^3$-2.09E-9 T$^4$+4.06E-13 T$^5$) *(1-1.5*(1-D)) |
| | 1473-1673 | 638*(1-1.5*(1-D)) |
| $\kappa$ (W.m$^{-1}$.K$^{-1}$) | 273-1673 | (1.96-2.32E-4 T+6.33E-7 T$^2$-1.91E-10 T$^3$)*(1-1.5*(1-D)) |
| $\rho$ (kg.m$^{-3}$) | 273-1673 | (6132 -9.23E-2 T-7.26E-5 T$^2$+4.58E-8 T$^3$-1.31E-11 T$^4$)*D |
| $\varepsilon'_r$ | 273-1673 | -5.38-4.34E-3 T+2.22E1 D+1.37E-2 T D |
| $\varepsilon''_r$ | 273-673 | 1.48E-1-5.76E-4 T-4.55E-01 D+1.77E-03 T D |
| | 673-873 | 3.82-6.03E-3 T-1.172E1 D+1.85E-2 T D |
| | 873-1073 | 1.56E1-1.95E-2 T-4.74E1 D+5.94E-2 T D |
| | 1073-1673 | 3.25E1-3.86E-2 T-7.64E1 D+8.46E-2 T D+3.82E-6 T$^2$+1.07 D$^2$ |

# Figures

Fig. 1: Zirconia permittivity as function of temperature and relative density calculated by the effective medium approximation: a) real part, b) imaginary part. *(one column wide)*

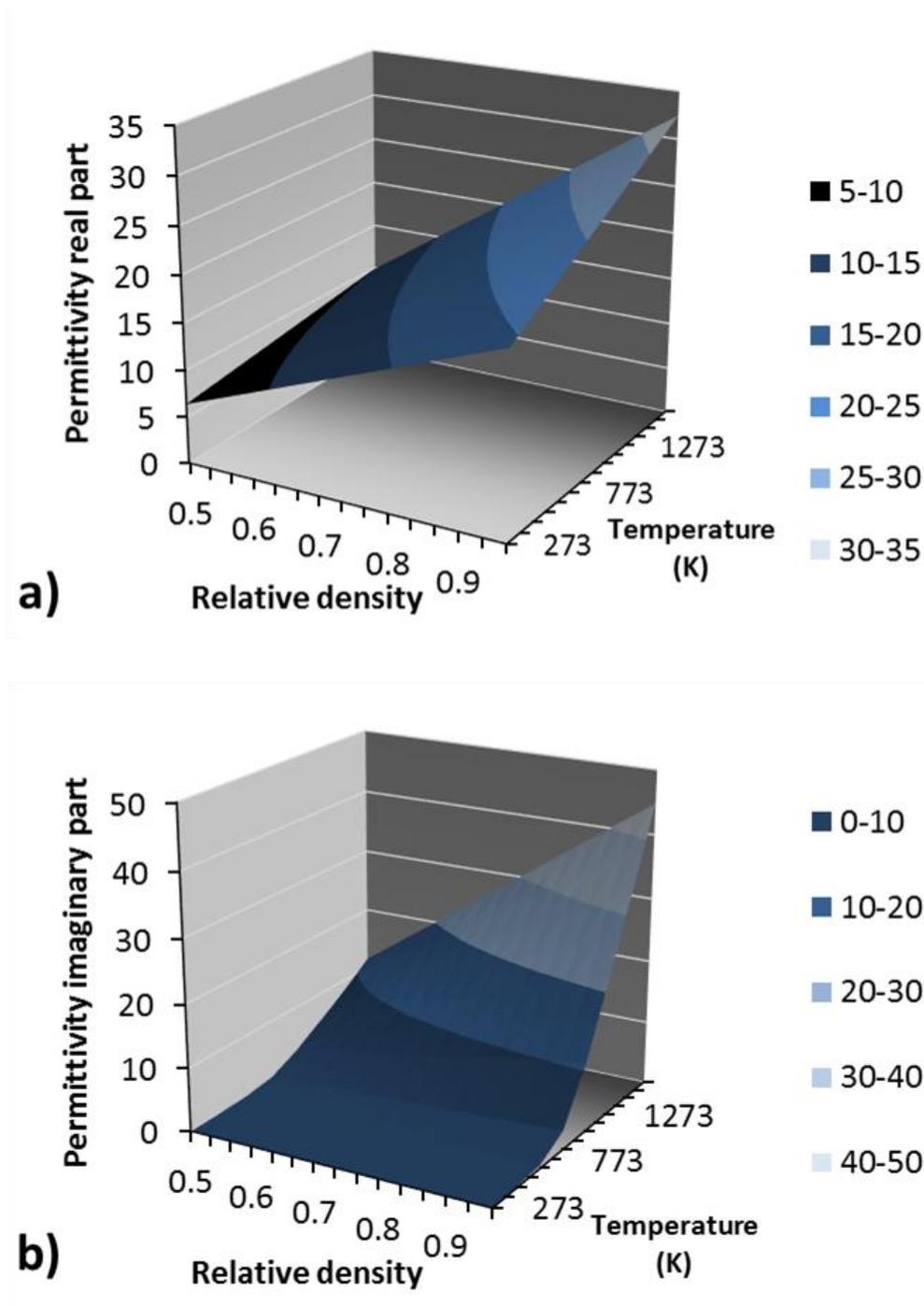

Fig. 2: a) Dimensions of the TE$_{102}$ resonant mode cavity, b) electric and magnetic field intensity inside the empty cavity, c) cavity with the zirconia sample.

*(one column wide)*

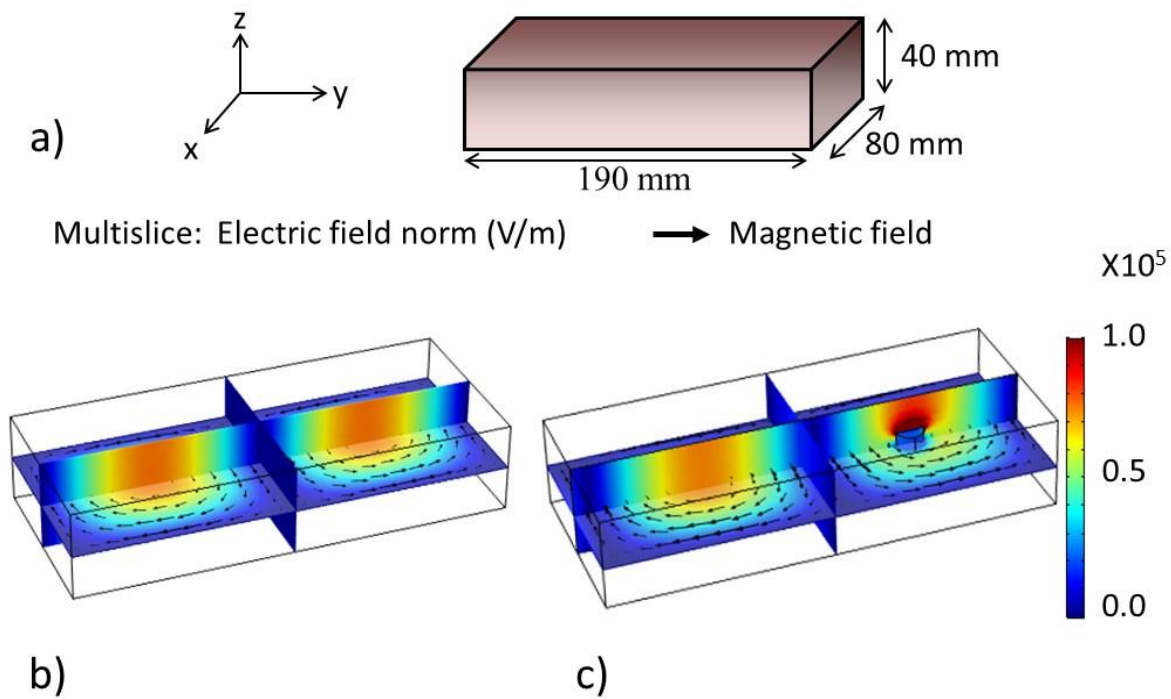

Fig. 3: Heating simulations of 3Y-ZrO2 at 55% of relative density: a) min, max, average temperature evolution for samples heated at different constant input powers; (b, c, d) sample temperature fields and electromagnetic losses for input power between 1000 and 2000 W at Tmax=1250 K (e, f, g). *(one column wide)*

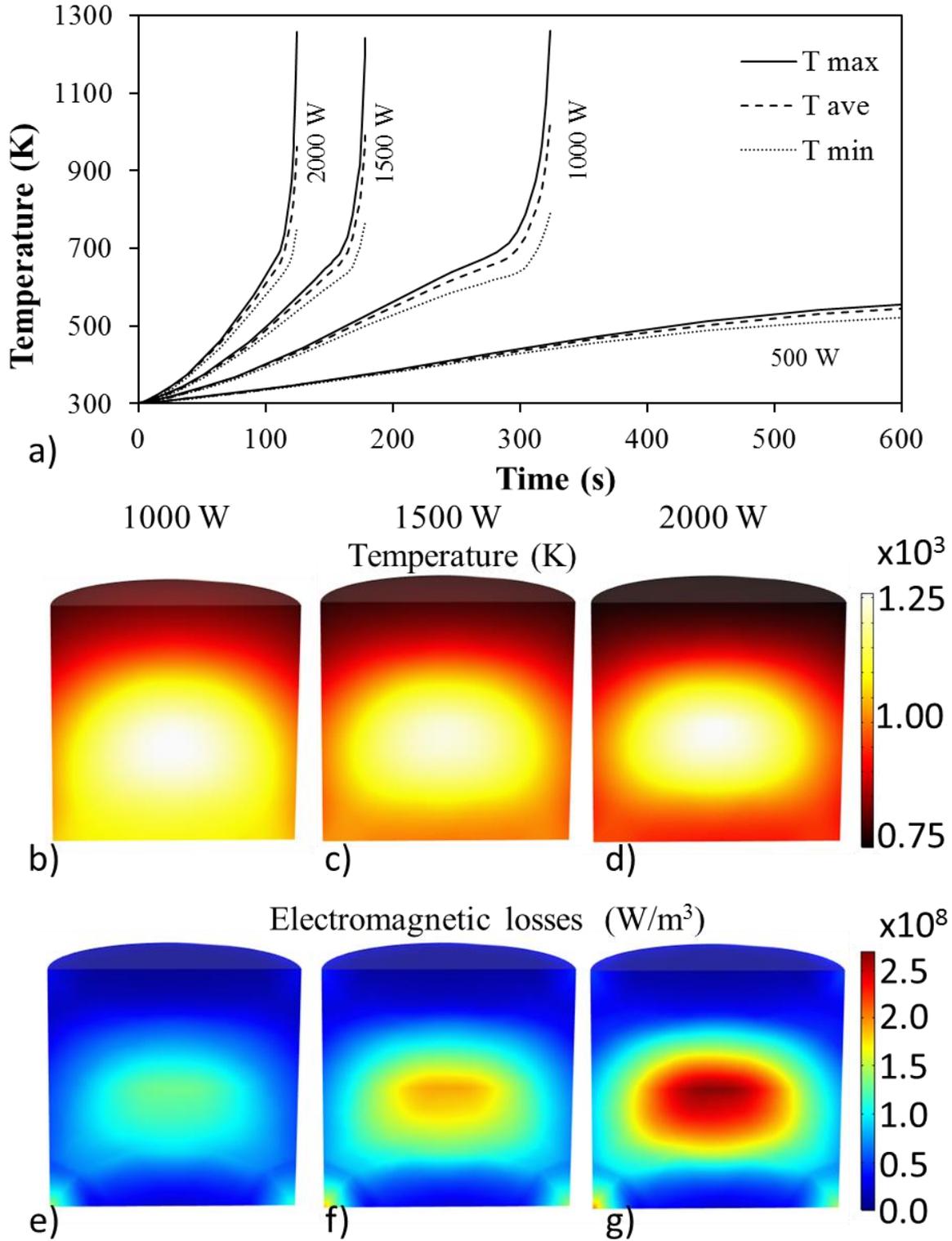

Fig. 4: Sample temperature fields at Tmax=1250 K for 2000 W: (a, b, c) with different sample locations, (d, e, f) with different sample dimensions. *(one column wide)*

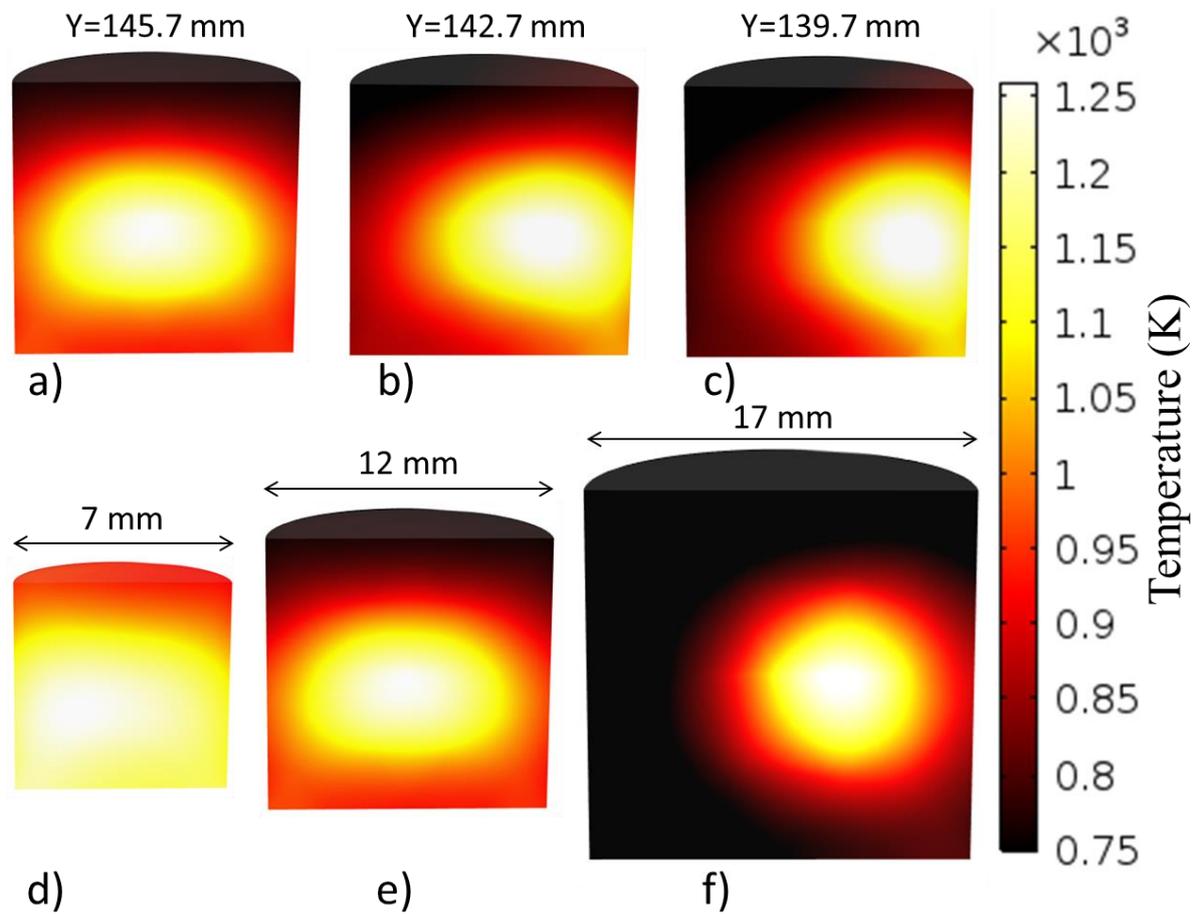

# Graphical Abstract

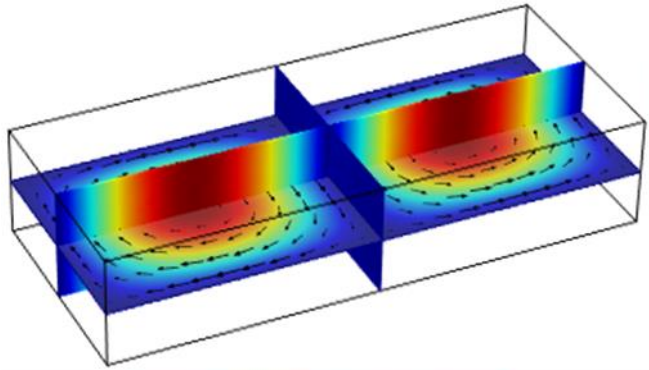
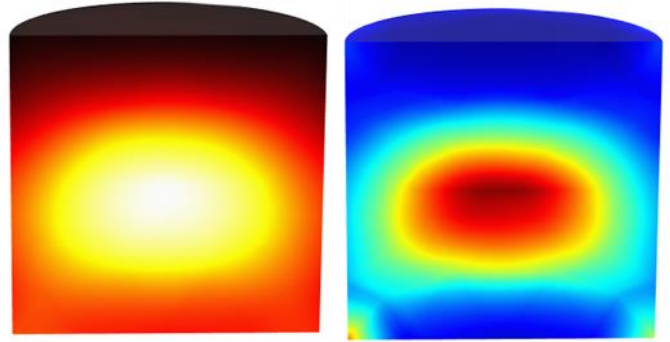
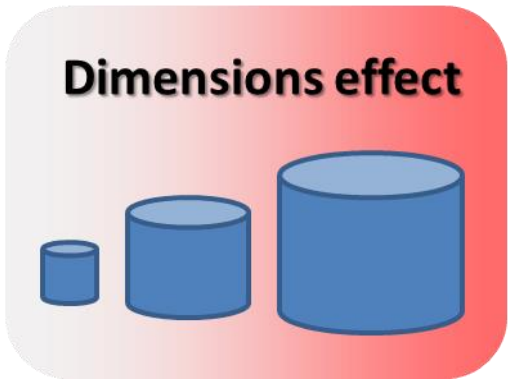
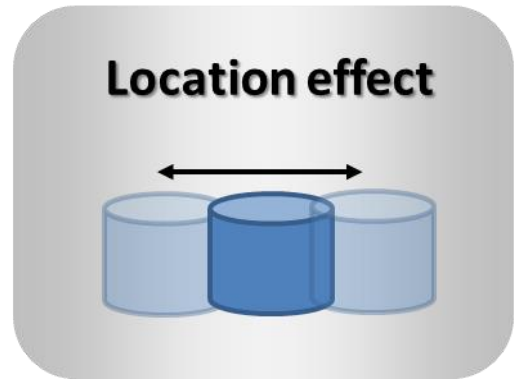